**Aqueous Laponite® Dispersions are Attractive Gels, Not Repulsive Wigner Glasses: A Critical Commentary**

Yogesh M Joshi,[1, a)] Shrajesh Patel,[1] and Khushboo Suman[2, b)]

[1]Department of Chemical Engineering, Indian Institute of Technology Kanpur, Kanpur 208016, India

[2]Department of Chemical and Biomolecular Engineering, University of Delaware, Newark, Delaware 19716, USA.

a) Author to whom correspondence should be addressed: joshi@iitk.ac.in

b) Electronic mail: ksuman@udel.edu

**Abstract:**

An aqueous dispersion of Laponite® has been studied in the literature for over the past three decades. Typically, the aqueous dispersion of Laponite® undergoes incessant evolution of its microstructure, wherein its elastic modulus and the mean relaxation time show a continuous increase as a function of time. A considerable amount of discussion has revolved around the nature of this dispersion, specifically whether it can be classified as a repulsive Wigner glass state, characterized by disconnected Laponite® particles stabilized by electrostatic repulsions, or an attractive gel state, in which the particles form a percolated space-spanning network. The proponents of the Wigner glass state also conjecture that this system experiences a glass-glass transition after a period of two days has elapsed since its preparation. In this commentary, we explore this topic from a rheological point of view analyzing the published literature and performing new experiments. Aided by additional evidence from the literature, we propose that rheological behavior overwhelmingly suggests that an aqueous dispersion of Laponite® undergoes sol - attractive gel transition and remains in the attractive gel state over at least up to 7 days without undergoing any additional transition. Importantly, rheology, despite being a macroscopic tool governed by principles of mechanics, offers profound insight into the microstructure of this particular system. The corresponding analysis conclusively determines the state of an aqueous dispersion of Laponite® to be an attractive gel.

An aqueous dispersion of Laponite® RD or XLG has fascinating physical properties and a myriad of commercial applications. Dispersion of a low concentration of Laponite® in water results in a dramatic increase in the viscosity and elasticity of the dispersion. As a result, an aqueous dispersion of Laponite® undergoes liquid to soft solid transition after a certain elapsed time, the rate of which strongly depends on the Laponite® and salt concentration. The understanding of the state behavior of an aqueous dispersion of Laponite® has a tumultuous history of three decades, with claims, counterclaims, and the same groups changing their positions over the years [1-4]. Various characterization tools have been used to study this system. On one hand, there is a consensus that Laponite® dispersion with a concentration between 1 and 2 weight percent forms an attractive gel. On the other hand, the major issue of contention has been whether spontaneously evolving aqueous dispersion of Laponite® above 2 weight % concentration without any externally added salt forms a repulsive Wigner glass or an attractive gel. Interestingly the school that proposes this system to form a Wigner glass also proposes it to undergo a so-called glass–glass transition [5, 6]. On the other hand, remarkably, despite being a bulk mechanical characterization tool, rheology has played a critical role in shaping this debate, particularly the validation of the Winter criterion that proposes this system to undergo sol-attractive gel transition [7-9]. In view of the continuing debate on Laponite® dispersion being an attractive gel or a repulsive glass and whether it undergoes glass–glass transition, the objective of this commentary is to deliberate on the state of aqueous Laponite® dispersion from a rheological perspective, substantiated by other indicators, and to put together a list of unambiguous arguments which confirm the Laponite® dispersion to be arrested in an attractive gel microstructure.

Before embarking on the discussion of microstructure, it is important to understand very briefly various terms and phenomena, such as a gel state, a glass state, and associated transitions. A Gel state is characterized by an infinite space spanning percolated network formed by one of its components. Furthermore, the characteristic lengthscale (mean distance between the two adjacent junctions) associated with the network is much larger than the lengthscale of the colloids [10]. In rheology, often the critical gel state is referred to, which is a unique state wherein the percolated network is least dense [11, 12]. This critical gel state or the point of gelation transition is observed to be independent of observation timescales (or frequencies) [12]. A glass state, on the other hand, originates from the caging effect due to the high concentration of one of its components. The characteristic lengthscale associated with the same, therefore, has the same order as interparticle distance [10]. The glass transition phenomenon is dependent on the observation timescale, and hence the point of glass transition depends on the probed frequency [13]. Typically, if $U$ is attractive interparticle energy, $U/k_BT$ ($k_B$ is Boltzmann constant and $T$ is temperature) represent relative intensity of the attractive interactions [14]. If $\phi$ is the volume fraction of the particulate matter, then for low values of $U/k_BT$ (attraction is not dominant) and $\phi$ one gets equilibrium liquid [10]. With an increase in $\phi$ but at low $U/k_BT$, the system undergoes glass transition. Furthermore, even at low values of $\phi$ but greater attraction between the particles (increase in $U/k_BT$) material forms an attractive gel state and at high values of both $\phi$ as well as $U/k_BT$, an attractive glass state can be obtained [10, 15]. Interestingly, gelation transition in particulate suspension may occur by crowding of clusters of particles rather than particles [14]. In such a case, it becomes difficult to distinguish to differentiate between some of the experimental signatures of a gel state in comparison with the glass state, though the above-mentioned distinction between the two states remains valid. This commentary discusses the microstructure of aqueous Laponite® dispersion, wherein the particles are anisotropic as well as possess opposite charges on the same. The density of Laponite® particles is 2.53 gm/cm$^3$, and the discussion in literature is confined to concentrations of the same below 4 weight % that is less than 2 volume % [4]. In addition, the electrostatic Debye screening length associated with faces of the particle also are needed to be considered. Consequently, all these aspects make the aqueous dispersion of Laponite® a unique case.

Introduction and brief review of the different views on the nature of Laponite® dispersions:

Laponite® is a synthetic smectite clay mineral. Numerous grades of Laponite® are commercially available, among which the RD and XLG grades are chemically the same but with different customer specifications [16]. In this commentary, henceforth, we will solely use the name Laponite® to refer to either the Laponite® RD or the Laponite® XLG. A primary particle of Laponite® has a disk-like shape with a diameter in the range of 25 to 30 nm and a thickness of 1 nm. Upon dispersing the same in the aqueous medium, it acquires a positive charge on its edge and a negative charge on its face. The physicochemical aspects associated with this phenomenon have been discussed in great detail elsewhere [4]. The dispersion eventually undergoes liquid - to - soft solid transition. The nature of the solid state that renders the dispersion non-zero elastic modulus over the explorable frequency (or observation time) range has been studied over the past 3 decades [1, 3, 8, 17, 18]. One school believes the aqueous dispersion of Laponite® above the concentration of 2 weight % without any externally added salt forms a repulsive Wigner glass [3, 18]. The other school, on the contrary, believes the aqueous dispersion of Laponite® forms an attractive gel state [1, 8]. In a large body of literature, Laponite® dispersion has been termed as Soft Glassy Material [19]. But it is important to note that the phrase "soft glassy materials" is highly broad and includes both repulsive glasses as well as attractive gels [20]. Therefore, it is important to differentiate between the states of the material being a repulsive Wigner glass or an attractive gel from the more general notion of soft glassy materials. As mentioned by Angelini *et al.* [6], repulsive Wigner glass is "an arrested state formed by disconnected particles stabilized by electrostatic repulsions." On the other hand, the attractive gel is a space-spanning percolated network of distinctly connected particles through attractive electrostatic and/or van der Waals interactions. In this commentary, we revisit the debate on the nature of microstructure in aqueous Laponite® dispersions by providing unambiguous rheological evidence of the existence of an attractive gel state. Furthermore, we aid the rheological inference with other supporting indicators. Finally, we address the claim in the literature that this system undergoes glass – glass transition by performing new experiments with carefully defined protocols.

Rheological evidence supporting the existence of attractive gel:

*Evidence from literature:*

The rheological behavior of aqueous dispersion of Laponite® above 2 wt. % with and without any salt is well documented in the literature [3, 7, 8, 21]. It has been observed that, in general, the elastic ($G'$) and the viscous ($G''$) moduli show the power law frequency dependence given by: $G' \sim \omega^p$ and $G'' \sim \omega^q$. These groups show that value of power law coefficients decrease (from an initial value of $p=2$ and $q=1$ associated with a liquid or sol state by definition) and reach such a state where $p=q$. Ultimately, the dispersion enters a soft solid state. While undergoing evolution, the state wherein $G'$ and $G''$ show identical power law dependence on frequency ($p=q$) is known to be critical gel state wherein the dissolved or dispersed component forms a space-spanning percolated network formed by attractive interaction (physical gels) or covalent bond formation (chemical gels) [12]. Interestingly, aqueous dispersion of Laponite® above 2 wt. % without any salt has been claimed to be a Wigner glass, to begin with, that undergoes glass – glass transition around 2 days after preparation [5]. In order to study the rheological behavior of this very system up to 7 days since preparation we perform standard oscillatory shear experiments in a controlled manner. We also carry out the ionic conductivity and pH measurements to support the rheological findings.

*New experimental evidence:*

We prepare 3 wt. % dispersion (same system as used by Angelini *et al.* [6]) by mixing oven dried (at 120°C for 4 h) Laponite® XLG in ultrapure water (pH 7 and resistivity 18.2 MΩ cm) using ultra turrax drive for 30 min. We use ARG2 (TA Instruments) rheometer with concentric cylinder geometry (cup diameter 30 mm and a gap of 1.2 mm) to perform the rheological experiments. Immediately after mixing is over, we filter the sample using 0.45 μm (Millex-HV, Sterile filter) filter directly into the shear cell. We cover the free surface of the Laponite® dispersion using low-viscosity silicon oil to prevent evaporation as well as free surface-induced anisotropic structure formation [22]. Subsequently, we subject the dispersion to oscillatory stress $\sigma = \sigma_0 \sin \omega t$ with $\sigma_0 = 0.1$ Pa. We also carry out strain-controlled experiment by applying $\gamma = \gamma_0 \sin \omega t$ with $\gamma_0 = 0.1$ %. Both these values of $\sigma_0$ and $\gamma_0$ are in the linear viscoelastic domain. The detailed experimental procedure including mutation numbers analysis is discussed elsewhere [8]. Each frequency ramp experiment takes 2 min. For the first day, we obtain dynamic moduli by applying the frequency ramp continuously. Beyond 1 day we apply a frequency ramp every 2 h. We continue the rheological experiments uninterrupted for 7 days without disturbing the sample. The temperature is maintained at 25°C. We perform the experiments in the stress-controlled mode two times while in the strain-controlled mode once for reproducibility. In addition to the rheological experiments, we also obtain the pH and ionic conductivity of the sample at regular intervals. The pH was determined using an F-71 pH meter with a conventional glass electrode, while the ionic conductivity was determined using a DS-71 conductivity meter with a 2-pole cell electrode (both instruments from HORIBA Scientific). The suspension was mechanically shear melted to turn it into a liquid, which allows for unrestricted ionic mobility, before measuring ionic conductivity and pH.

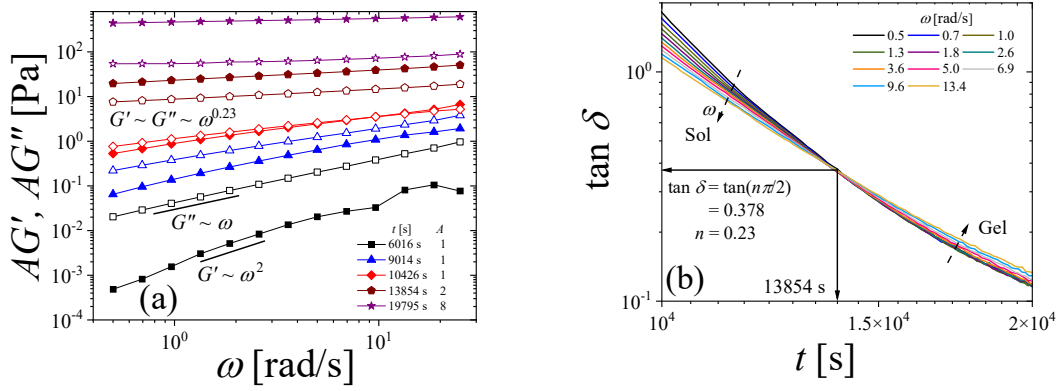

Figure 1. (a) The elastic ($G'$, filled symbols), and viscous ($G''$, open symbols) moduli are plotted as a function of frequency at different times elapsed since the preparation of the sample. (b) Iso-frequency $\tan\delta$ curves are plotted as a function of time elapsed since the preparation of the sample. The behavior shows a clear rheological signature of sol–gel transition (refer to the text for more details). It should be noted that the sample was not disturbed after the freshly prepared filtered sample was loaded into the shear cell. To prevent evaporation losses and the formation of anisotropic structures, the free surface was covered with low-viscosity silicone oil over the entire duration.

Figure 1(a) depicts the variation of $G'$ and $G''$ with respect to $\omega$ for different values of waiting time $(t_w)$ up to $t_w \cong 20,000$ s. During the initial stages, $G'$ and $G''$ exhibit characteristics of a liquid, wherein $G' \sim \omega^2$ and $G'' \sim \omega$. As time progresses, the $\omega$ dependence of both the moduli becomes progressively less pronounced, and at the critical gel state, both moduli exhibit a similar power law relationship on $\omega$ given by [12]:

$$G'' = G' \tan\left(\frac{n\pi}{2}\right) = \frac{\pi S}{2\Gamma(n)\cos(n\pi/2)}\omega^n, \qquad (1)$$

where $n$ is the critical relaxation exponent, $S$ is the gel strength, and $\Gamma(n)$ is the Euler gamma function of $n$. Accordingly, from Figure 1(a), we get the critical relaxation exponent $n = 0.23$. In Figure 1(b), we plot the temporal evolution of $\tan\delta$ (which is $G''/G'$) curves at various $\omega$, which demonstrate a decreasing trend with an increase in $t_w$. Furthermore, it is observed that the rate of decrease in $\tan\delta$ weakens as $\omega$ increases. It is noteworthy that all the iso-frequency curves of $\tan\delta$ intersect at a common point, referred to as the critical gel state. The value $\tan\delta$ at the cross-over point is independent of $\omega$, and is given by $G''/G' = \tan(n\pi/2)$, which also yields a value of $n = 0.23$. Importantly this value of $n$ obtained from the $\tan\delta$ crossover is found to be identical to the value obtained from power law dependence of dynamic moduli on frequency thereby confirming internal consistency.

The dependence of dynamic moduli on frequency given by Eq. 1 and its experimental validation shown in Figure 1 (at time 13854 s ~ 3.8 h) is of profound significance in the field of rheology and is known as the Winter criterion [12]. This unique state is known as the critical gel state, wherein Laponite® particles form percolated space-spanning network through attractive interactions in the dispersion. Validation of the Winter criterion with the critical power law coefficient $n$ has many implications as mentioned below:

1. If we consider the degree of crosslinking to be $p$, whose value at the critical gel state is $p_c$, then it is known that the zero shear viscosity ($\eta_0$) and the equilibrium modulus ($G_e$) show the dependence given by: $\eta_0 \sim (p_c - p)^{-s}$ for $p < p_c$ and $G_e \sim (p - p_c)^z$ for $p > p_c$, where exponents $s$ and $z$ are related to each other through $n$ as: $n = z/(z+s)$ [11]. This relation has recently been verified for the Laponite® dispersion in the vicinity of the critical state [11]. Furthermore, in the pre-gel state ($p < p_c$), according to de Gennes [23], size of largest aggregate $(\xi)$ grows as: $\xi \sim (p_c - p)^{1-z}$.

2. For chemically and physically crosslinking systems, in the vicinity of the critical gel state, the evolution of dynamic moduli exhibits a power-law dependence given by [12]:

$$\left(\frac{\partial \ln G'}{\partial p}\right)_{p=p_c} = C\left(\frac{\partial \ln G''}{\partial p}\right)_{p=p_c} \sim \omega^{-\kappa}, \qquad (2)$$

where $\kappa$ is the dynamic critical exponent and $C$ is the proportionality constant. In Eq. (2), since $|t - t_c| \sim |p - p_c|$, the differentiation is carried out with respect to $t$ and is obtained at $t = t_c$. This relation has been validated for the Laponite® dispersion in the vicinity of the critical gel state [8, 11]. We also calculate these parameters for the current system under investigation and obtain $\kappa = 0.25 \pm 0.01$ and $C = 2.47 \pm 0.48$. Very interestingly, as a consequence of symmetric divergence of longest relaxation scale in the vicinity of critical gel

state relates $\kappa$ to $n,s$ and $z$ as: $z = n/\kappa$ and $s + z = 1/\kappa$ [11]. Importantly, for a very broad range of gel forming materials that include chemically and physical crosslinked polymers (that include over 30 experiments on the Polydimethylsiloxane system with varying stoichiometry, concentration, and precursor chain length) [12, 24] and colloidal particles [11], $\kappa$ has been observed to lie between 0.15 and 0.25 while $C$ has been observed to be close to 2 [8, 25]. This observation puts aqueous dispersion of Laponite® in the broad family of gel forming materials that show similar behavior.

3. The critical gel state has a fractal structure, and assuming a complete screening of the excluded volume effects, it leads to a fractal dimension ($f_d$) of the same given by: $f_d = 5(2n-3)/2(n-3)$ [26]. Interestingly, this finding has been verified for a variety of other systems by performing rheology and scattering experiments at the same time [27-29].

4. At the critical gel state, the relaxation time spectrum $H(\tau)$ depends on relaxation time $(\tau)$ as per $H(\tau) \sim \tau^{-n}$ [12]. This suggests that Laponite® dispersion passes through a state where the high relaxation modes are sparsely populated while small relaxation modes are densely populated. According to Winter [30], the negative power law dependence of $H(\tau)$ on $\tau$ is the most significant rheological characteristic feature of an attractive gel state. Conversely, Winter [30] emphasized that dependence of $H(\tau)$ on $\tau$ has a positive exponent for glasses. Recently Suman and coworkers [31] discussed not just the features of $H(\tau)$ at the critical gel state, but also how the shape of relaxation time spectra transforms from the sol to the post-critical gel state in the context of Laponite® dispersion.

Strictly from a rheological perspective, the above discussion clearly suggests that the studied 3 wt. % aqueous dispersion of Laponite® shows all the characteristic features of an attractive colloidal gel. The last point also clearly indicates how the dependence $H(\tau) \sim \tau^{-n}$ cannot lead to a repulsive glass [30]. Furthermore, rheologically, a material undergoing liquid-to-glass transition is known to show a maximum in $\tan\delta$ whose peak value shifts with a change in frequency, as the glass transition phenomenon depends on observation timescale [13, 32]. On the contrary, on one hand, the present system does not show any maximum in $\tan\delta$. On the other hand, at the critical gel state, $\tan\delta$ is observed to be independent of frequency that suggests a time invariant state associated with the gelation transition. As per the phase diagram proposed by Ruzicka and Zaccarelli [3], Laponite® dispersion forms respectively an attractive gel and repulsive glass below and above 2 wt. % concentration. However, Jatav and Joshi [8] confirm the validation of the Winter criterion for the entire spectrum of concentration of Laponite® (1.4 to 4 wt. %) and NaCl concentrations (0 to 7 mM). On the other hand, Cocard *et al.* [9] validate the Winter criterion for 1 wt % using conventional rheometry while Rich *et al.* [33] validate Winter criterion for 1 wt %+ 5.9 mM NaCl dispersions using microrheology. A complete list of Winter criterion validation by aqueous Laponite® dispersion below and above 2 wt. % through the conventional as well as microrheology is given elsewhere [4]. This clearly suggests the state of Laponite® below and above 2 wt. % cannot be different and the whole spectrum of concentrations shows an attractive gel state.

We also obtain the ionic conductivity and pH of the dispersion for over 7 days after preparation of the same. We observe that while both the parameters show a slight increase over 7 days, the change is less than 4 %. The mean value of the measured ionic conductivity is $\sigma = 1071$ μS/cm while the mean pH is given by: 9.9. The fact that the pH of ultrapure water (that is 7) increases after incorporation of Laponite®, there must be an increase in OH⁻ ion concentration. As explained elsewhere [4, 34], this increase in OH⁻ ion concentration is attributed to the protonation of the Laponite® particle edges, which causes it to acquire a positive charge. Furthermore, knowledge of ionic conductivity leads to the Debye–Hückel screening length $\left(\lambda_D\right)$, whose mean value is estimated to be: $\lambda_D \approx 2.9$ nm. The detailed calculations on how to obtain $\lambda_D$ is given by Jatav and Joshi [35]. This discussion clearly suggests that electrostatic Debye screening length has the same order of magnitude as thickness of the Laponite® particle (≈ 1 nm) and is typically an order of magnitude smaller than the particle diameter (≈25 nm). Consequently, with a positive edge charge, and significantly small $\lambda_D$ that suggests length-scale associated with repulsive interaction, it is extremely difficult to perceive that particles remain disconnected and self-suspended owing to face–to–face repulsion, which is necessary for the Wigner glass formation.

Complementary experimental evidence:

In the literature, other than rheology, various methodologies, such as dilution, simulations, SAXS, SANS, the effect of pH, effect of externally added ingredients (such as salt, sodium pyrophosphate, Polyethylene oxide, and Pluronic), microscopy, etc., have been employed to study the microstructure of aqueous dispersion of Laponite®. Among these, the dilution study by itself has been observed to be inadequate to corroborate the nature of the interactions (attractive/repulsive) and the microstructure of the dispersion [1, 36]. The simulations performed on this system, on the other hand, are highly dependent on a variety of factors pertaining to the system such as the number density, Debye screening length, exact value of charges on the edges and faces, size and the aspect ratio of particles, etc. used in the simulations, as well as how closely the underlying assumptions correspond to the actual nature of the system [4]. As a result, simulations do not provide a definitive representation of the microstructure. However, several studies do provide distinct evidence supporting the existence of a sol-attractive gel transition in an aqueous Laponite® dispersion, which we have presented in Table 1. All the mentioned observations, namely, accelerated liquid-solid transition with an increase in salt concentration and corresponding DLVO analysis, retarded liquid-solid transition with the addition of sodium pyrophosphate only below the pH of isoelectric point, validation of the Winter criterion, photogelation upon addition of acid and verification by SANS and Cryo-TEM images showing edge to face bond overwhelmingly point towards the presence of attractive gel state in this system. These observations, particularly the fact that the increase in elastic modulus gets accelerated with the addition of salt, conclusively suggest that the incipient structure cannot be dominated by repulsive interactions wherein particles remain disconnected from each other.

Table 1. List of evidence that supports the origin of microstructure to be attractive gel in Laponite® dispersion.

| 1 | Accelerated liquid-solid transition with an increase in salt concentration and DLVO theory. [8, 36-40] |
|---|---|
| | The addition of salt mitigates the repulsive potential, which accelerates the formation of structures. Such accelerated gelation suggests the system to be in an attractive gel state. The observations of the DLVO analysis confirm that the rate of structure formation accelerates with an increase in salt concentration, as this causes a decrease in $\lambda_D$. If repulsive interactions were responsible for the |

|   | structure formation (as they do in the Wigner glass), the structure would have formed more slowly due to the addition of salt. |
|---|---|
| 2 | Retarded liquid-solid transition with the addition of sodium pyrophosphate [1, 36, 41-43]<br><br>Dissociation of sodium pyrophosphate in aqueous media results in the formation of tetravalent anions, which attach to the positively charged edges of Laponite®. As a result, there is a reduction of edge-face interaction thereby delaying the gelation (liquid-solid transition) process. At higher concentrations of sodium pyrophosphate, the gel completely transforms into liquid. |
| 3 | Effect of pH and sodium pyrophosphate [36]<br><br>It is known that for the pH above the isoelectric point ($\approx 11$) the edge of the Laponite® particle acquires a negative charge and the Laponite® particles aggregate through the van der Waals interactions. Interestingly, while sodium pyrophosphate retards the gelation of Laponite® dispersion below the pH of 11, at pH 13 sodium pyrophosphate showed no effect on Laponite® dispersion. These results very clearly suggest microstructure formation in Laponite® dispersion below pH of 11 is through the edge-to-face attractive interaction and no Wigner glass can be present. |
| 4 | Rheology<br><br>Numerous rheological studies that show validation of the Winter criterion conclusively suggest the presence of rheological sol-gel transition in aqueous Laponite® dispersion [7-9, 21, 33]. The detailed scaling analysis of the critical exponents [11], the corresponding relaxation spectra in the pre and post gel state [31], the convolution relation and the Kramers–Kronig relations in a limit of linear viscoelasticity [44], and the non-linear viscoelastic behavior [44] of this system show all the signatures that have been reported for polymeric systems undergoing chemical gelation [12]. |
| 5 | Photogelation upon addition of acid and verification by SANS [45]<br>The stable sol Laponite® dispersion (with stabilizing nonionic surfactant such as Polyethyleneoxide, and pluronic) got converted to gel upon addition of a photoacid generator (PAG). Upon shining light, PAG undergoes photolysis, lowering pH by $\approx 3$ units. This displaces stabilizing surfactant from nanoparticle disk faces, facilitating a three-dimensional "house-of-cards" network, spanning the entire sample volume. The presence of fractal network structure is validated |

| | by an upturn in the intensity in a limit of small values of wave vector $q$ in Small Angle Neutron Scattering (SANS) experiments. Again, the photogelation results are consistent with the attractive gel structure rather than the repulsive Wigner glass. |
|---|---|
| 6 | Microscopy [4, 8]<br>    Cryo-TEM images of Laponite® dispersion shows homogeneously distributed particles engaged in a space-spanning percolated network with edge-to-face attractive bonds. |

Comment on the proposed glass-glass transition:

We will now explore the scientific evidence for the existence of a glass-glass transition. The comprehensive reasons why Laponite® dispersion does not arrest in a repulsive glass state have been thoroughly addressed in our previous work [4]. Recently, there have been reports on glass-glass transition in an aqueous dispersion of Laponite® having a concentration of 3 weight % without any externally added salt [5, 6]. Their primary proposal is that over the initial period of around 2 days after preparation, an aqueous dispersion of Laponite® forms a repulsive Wigner glass. Subsequently, over a period of one day, it undergoes a transition to a second glassy state which they call disconnected House of Cards. The glass-glass transition is characterized using dynamic mechanical analyzer (DMA), rheology and X-Ray photon correlation spectroscopy (XPCS). It is reported that Young's (compressive) modulus ($E$), elastic modulus in oscillatory compression ($E'$), residual strain ($\varepsilon_r$) and elastic modulus in oscillatory shear ($G'$) are practically same for measurements carried out before day 2. On day 3, these parameters show a sudden jump [5, 6]. The relaxation time $\tau$ is also reported to show anomalous behavior where it increases for about 1.5 days, followed by a decline lasting slightly over 2 days, and then undergoes a substantial increase, reaching just over 3 days [5, 6]. While the abrupt changes in the mechanical parameters have been associated with glass-glass transition, it is important to note that the variation in parameters do not explicitly imply of glass–glass transition.

Motivated by the distinctive ability of rheology to distinguish between gel and glass states, we aim to employ rheology to investigate the evolution of dynamic moduli over a duration of 7 days. In order to check the abrupt jump in elastic modulus after 2 days as reported, we now explore how elastic and viscous moduli evolve over a period of 7 days. Figure 2(a) displays the plots of $G'$ and $G''$ against the waiting time ($t_w$) at $\omega =10$ rad/s. As demonstrated, initially $G' \ll G''$ suggesting suspension to be in a liquid-like state. With increases in $t_w$, both moduli undergo an increase with $G'$ exhibiting a higher rate compared to $G''$, leading to a crossover of $G'$ and $G''$. Beyond the cross over, $G''$ shows a weak decrease with significant scatter in the data; $G'$ on the other hand, changes slope and continues to increase with a power law exponent dependence of $G' \sim t_w^{0.56}$ over the duration from 10 hr up to 7 days. In Figure 2(b) we plot $G'$ as a function of $\omega$ up to 7 days at an interval of 1 day. It can be seen that, $G'$ is independent of explored $\omega$. In addition, $G'$ increases gradually from day 1 to 7 without showing any abrupt change at any time. The observed continuous steady increase in $G'$ has been attributed to the densification of the network and strengthening of the edge-to-face bonds as a function of time. The above discussion is strictly for unrejuvenated (spontaneously evolving) Laponite® dispersion. How rejuvenation affects the attractive gel network of Laponite® particles is discussed elsewhere [46].

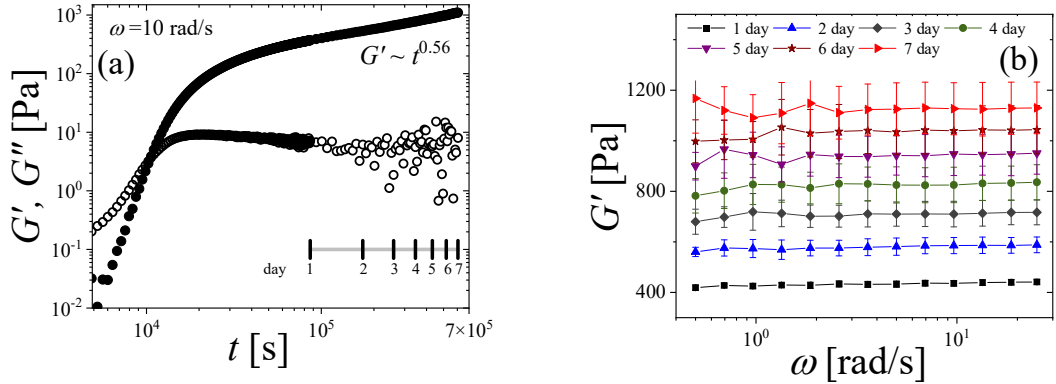

Figure 2. (a) The time evolution of elastic ($G'$, filled symbols), viscous ($G''$, open symbols) moduli are plotted at a single frequency for the same experiment whose data is shown in Fig. 1. Positions associated with 1 to 7 days related to horizontal axis have been marked on the plot for a ready reference. (b) The elastic modulus is plotted as a function of frequency from day 1 to 7 at an interval of 1 day. The mean value and error bar are obtained by analyzing three experiments. It should be noted that the sample was not disturbed for the entire duration of 7 days after the freshly prepared filtered sample was loaded into the shear cell. To prevent evaporation losses and the formation of anisotropic structures, the free surface was covered with low-viscosity silicon oil over the entire duration.

We shall now discuss the additional supporting evidence present in the literature [5] to claim the presence of repulsive Wigner glass and glass–glass transition in this system as follows:

1. Angelini *et al.* claim that while studying dynamic correlation function through XPCS, rejuvenated samples show a crossover from a stretched to a compressed exponential occurring around three days after rejuvenation [5]. Firstly, it is not clear why this observation has to be exclusive to the transition from Wigner glass. However, it is important to note that in the recent work of Angelini *et al.* [6], the non-rejuvenated samples do not show any such transition (although they do report fluctuation). It is therefore not clear whether this test any longer holds to claim so-called glass–glass transition, and if yes what is the microstructural reasoning to exclude attractive gels from the same. On similar lines, they observe a dip in relaxation time (measured at $Q = 0.1$ nm$^{-1}$) between 1.5 and 2 days as it evolves with time. However, it is not clear why such a dip cannot be seen in their sample if it is composed of attractive gel. It would indeed be interesting to investigate the microstructural origin of such a dip, but prima facie there is nothing that makes this dip limited only to a transition from Wigner glass.
2. It has also been reported that over a similar range of time (2 to 3 days) after rejuvenation, in the Small Angle X-ray Scattering (SAXS) experiment the peak of the structure factor $S(Q)$ shifts its position by about 10% [5]. It is important to note that Laponite® is an anisotropic particle with an aspect ratio of around 25 and it has anisotropic charge distribution. Consequently, in the dispersion, its orientation is not independent of its position. As discussed by Greene *et al.* [47], Nicolai and Cocard [48], and Suman and Joshi [4] in detail, under such circumstances the decoupling approximation employed to obtain $S(Q)$ for Laponite® dispersion can result in spurious peaks, which do not convey any information about the microstructure.

Effectively, in our opinion, the foregoing discussion establishes the lack of empirical evidence for the presence of Wigner glass as well as the so-called glass-glass transition within an aqueous dispersion of Laponite®. Overall, the above discussion shows that the studied 3 wt. % Laponite® dispersion does not show any rheological signatures of repulsive Wigner glass, nor does it show any transition over 7 days under undisturbed and controlled conditions.

## Outlook:

It is important to note that Laponite® clay used in most of the studies over the past 3 decades has been produced on an industrial scale and may have batch-to-batch variation concerning different levels of salt present in the same. In any scientific study, it is, therefore, important to report ionic conductivity and pH of water as well as the aqueous dispersion of Laponite®. Furthermore, Laponite® dispersion has also been observed to be sensitive to the sample preparation protocol including removing adsorbed water by baking the 'dry' Laponite® powder, filter size used for filtration of sample before initiating experiments, adjusting pH of solution, using low-viscosity oil to prevent evaporation and addition of any salt which alters the ionic concentration [2]. In addition, the Laponite® dispersion is extremely fragile in the initial days after preparation. Consequently, any movement of the same may cause breakage of the interparticle bonds and hence extreme care needed to be exercised. Moreover, it is also important to take into account macroscopic anisotropic structures that get formed in Laponite® dispersion when any of its (free) surfaces get exposed to air [22]. All these aspects profoundly affect the physical behavior of Laponite® including its rheological behavior [4]. Nonetheless for the reasons discussed above, irrespective of the batch – to batch variation in production and/or variance in the preparation protocol, we believe that the proposal of a repulsive Wigner glass structure in an aqueous dispersion of Laponite® is untenable.

To summarize, in this commentary, we discuss the possibility of having the Wigner glass state and glass-glass transition in an aqueous Laponite® dispersion. However, through rheology and other indicators, we observe that this system undergoes sol–gel transition. In addition, in our oscillatory shear experiments on a conventional rheometer that have been performed in a controlled fashion, we do not observe any transition in the rheological behavior up to 7 days. Finally, it is heartening to note that even though rheology is a bulk tool governed by the laws of mechanics, it provides keen insight into the microstructure of this system, which establishes the aqueous dispersion of Laponite® as an attractive gel.

**Acknowledgments:** We acknowledge financial support from the Science and Engineering Research Board, Government of India (Grant No. CRG/2022/004868).

**Data Availability Statement:** The data that support the findings of this study are available from the corresponding author upon reasonable request.